\documentclass{ws-procs975x65}

\begin{document}

\title{
BLACK HOLES IN HIGHER DIMENSIONS\\
(BLACK STRINGS AND BLACK RINGS)
}

\author{JUTTA KUNZ}

\address{Institut f\"ur Physik, Universit\"at Oldenburg, Postfach 2503\\
D-26111 Oldenburg, Germany\\
E-mail: jutta.kunz@uni-oldenburg.de}

\begin{abstract}
The main focus of this session 
was the presentation of new higher-dimensional black hole solutions,
including black rings, black strings, and multi black holes,
and the study of their properties.
Besides new asymptotically flat 
and locally asymptotically flat black objects 
also new black holes with anti-de Sitter asymptotics
were reported.
The studies of their properties included the investigation
of their stability,
their thermodynamics, their analyticity and their existence.
Furthermore, the geodesics in such higher-dimensional
space-times were investigated.
\end{abstract}


\bodymatter

\section{Introduction}\label{intro}

Today string theory plays a major role as a candidate 
for the quantum theory of gravity 
and the unification of all interactions. 
String theory 
requires higher dimensions for its mathematical consistency,
which presents a strong motivation to study the possible physical consequences
of the presence of more than four space-time dimensions.
One very active area here is the study of higher-dimensional
black holes, where
the extra dimensions are considered to be either infinite or compact.

Moreover,
for many years now the Anti-de Sitter/conformal field theory (AdS/CFT)
correspondence has been employed  as a means to study 
strongly coupled theories, where perturbative methods are not applicable.
This correspondence  allows to
translate dynamical problems in strongly coupled theories
in $D$ dimensions
into gravitational ones in $D+1$ dimensions.
This implies that finding new black holes and other black objects
in $D+1$ dimensions
allows for the prediction of new thermal phases in the dual field theories
in $D$ dimensions.

In four dimensions the stationary,
asymptotically flat electrovac black holes are given by the 
family of Kerr-Newman solutions.
The Kerr-Newman black holes are uniquely characterized by their
global charges, the
mass, the angular momentum and the electromagnetic charge(s).
A spatial section of their event horizon
has the topology of a two-sphere, $S^2$,
other horizon topologies are not allowed.

Already in 1986 Myers and Perry
\cite{Myers:1986un} obtained the generalizations of the
four-dimensional Kerr solutions.
These Myers-Perry black holes represent 
asymptotically flat rotating vacuum black holes 
in $D$ dimensions, which possess a spherical horizon topology, $S^{D-2}$.
Moreover, Myers and Perry argued that also black holes
with a ring topology of the horizon should exist.
When the five-dimensional vacuum black ring with horizon topology
$S^1\times S^2$ was found by Emparan and Reall \cite{Emparan:2001wn},
this created a surge of interest.
Soon many related black objects were found in five dimensions
including black saturns,
black di-rings or bicycling black rings
(for a review see e.g.~\cite{Emparan:2008eg}).

\section{New higher-dimensional solutions}

In the following we will discuss the various types of new solutions,
that were presented in the session,
starting with black rings in six dimensions.

\subsection{Black rings in more than five dimensions}

While there are a number of construction methods known to obtain
exact black hole solutions in five dimensions, there does not
seem to exist a general analytic framework for
the construction of black objects in more than five dimensions.
Therefore one has to either resort to perturbative methods
or to numerical methods at the moment.

Myers and Perry already suggested a
heuristic way to construct black rings,
namely to take a Schwarzschild black string,
to bend it, and to
achieve balance by spinning it along
the ring $S^1$ direction \cite{Emparan:2001wn}.
The perturbative technique of matched asymptotic expansions
\cite{Emparan:2007wm,Emparan:2009vd}
and blackfolds \cite{Emparan:2009cs}
is based on this picture.

The blackfold approach has led to perturbative solutions
for black rings in more than five dimensions,
when the ring radius is sufficiently large \cite{Emparan:2007wm}.
However, it cannot deal with solutions,
whose ring radius is not large
as compared to the radius of the $S^{D-3}$-sphere.
Thus, to obtain the phase diagram of higher-dimensional
black rings, numerical methods are called for.

In his talk Eugen Radu
presented solutions for balanced black rings in six dimensions,
that were obtained numerically,
employing two different methods, 
a finite difference solver and a spectral method
\cite{Kleihaus:2012xh}.
The general picture he unveiled for these black rings
exhibits remarkable similarities to the five-dimensional case.
In particular, there are again two branches of black ring solutions.
The scaled horizon area $a_{\rm H}$ possesses a cusp
at a minimal value of the scaled angular momentum $j$,
where it assumes its maximal value.
Apart from the region close to the cusp, 
the thin black ring branch is remarkably well
approximated by the blackfold approach.

Starting from the cusp, the branch of fat black rings
extends only over a short interval of the scaled angular momentum.
In this range there exist three different solutions 
with the same global charges, violating uniqueness
as in five dimensions.
This short branch is expected to 
end in a critical merger configuration
\cite{Emparan:2007wm,Emparan:2010sx}. 
Here a branch of `pinched' black holes should be approached in a
horizon topology changing transition.
The branch of `pinched' black holes itself
is expected to branch off from a critical Myers-Perry solution,
where a Gregory-Laflamme-like instability arises
\cite{Gregory:1993vy,Emparan:2003sy,Dias:2009iu,Dias:2010eu,Dias:2010maa,Dias:2011jg}.
It remains a numerical challenge
to obtain the branch of `pinched' black holes.

Eugen Radu concluded his talk with a discussion of other black hole solutions, 
amenable to these numerical methods.
In particular, he discussed static solutions with
horizon topology $S^2 \times S^{D-4}$ obtained
previously \cite{Kleihaus:2009wh}
and static composite black objects in more than five dimensions
\cite{Kleihaus:2010pr}.
While these static solutions are unbalanced,
the inclusion of rotation or of higher curvature terms
might balance them.
Recently, balance of $S^2 \times S^{D-4}$ solutions
could be achieved with a
help of an external magnetic field \cite{Kleihaus:2013zpa}.
Thus these solutions
approach asymptotically a Melvin universe background.

\subsection{Anti-de Sitter black holes and solitons
with scalar hair}

Let us now turn to higher-dimensional black holes in the framework of
the AdS/CFT correspondence \cite{Maldacena:1997re}.
A main point of interest here has been the description  of so-called
holographic superconductors with the help of black holes
in higher dimensional space-times
\cite{Gubser:2008px,Hartnoll:2008vx,Hartnoll:2008kx,Horowitz:2008bn}.

As observed by Gubser \cite{Gubser:2008px},
below a critical temperature
electrically charged black holes can become unstable to the formation
of scalar hair.
This instability results from the fact, that
the effective mass of the scalar field can
assume negative values below the
Breitenlohner-Freedman (BF) bound \cite{Breitenlohner:1982bm,Breitenlohner:1982jf}
close to the black hole horizon.
The scalar field then assumes a finite value on the horizon.
Its value on the AdS boundary is then associated with
a condensate of the boundary field theory. 
The reason for this behaviour is related 
to the fact that the scalar field possesses an effective mass 
that depends on the geometry of the
space-time if the scalar field is charged under a U(1). 
Close to the horizon the effective mass
can drop below the BF bound, 
while asymptotically it stays well above, 
ensuring the stability of asymptotic AdS.

An interesting open issue here was the question
whether higher curvature corrections can suppress the condensation.
In \cite{Gregory:2009fj} it was shown, 
that the formation of scalar hair cannot be suppressed
in Einstein-Gauss-Bonnet gravity in the ``probe limit'',
where backreaction is neglected.
In her talk Betti Hartmann addressed the problem
away from the probe limit, based on the self-consistent
treatment of the coupled system of Einstein-Gauss-Bonnet-matter equations
\cite{Brihaye:2010mr}.
  %``Holographic Superconductors in 3+1 dimensions away from the probe limit,''

In her study of the formation of scalar
hair on electrically charged black holes in five dimensional
anti-de Sitter space-time, the Gauss-Bonnet coupling constant
was treated as a parameter.
The results, however, showed, that - analogous to the probe limit -
a critical temperature exists, where condensation sets in.
While this critical temperature decreases,
as the Gauss-Bonnet coupling constant is increased,
it remains positive for any value of this coupling 
and for arbitrarily large gravitational coupling
\cite{Brihaye:2010mr}.

While her first study \cite{Brihaye:2010mr} considered planar
black holes and a charged scalar field, 
she subsequently investigated
hyperbolic black holes with a neutral scalar field
\cite{Brihaye:2011hm}.
Also for these hyperbolic black holes in Einstein-Gauss-Bonnet gravity
it was found that within a certain range of the scalar field mass 
they become unstable to the condensation of the scalar field. 
Interestingly, a whole family of 
hairy black hole solutions was found, that could be labeled by the number of
nodes of the scalar field function.

In the case of an uncharged scalar field it is not the coupling 
to the U(1) gauge field that leads to the condensation, 
but rather the fact that the near-horizon geometry
of a black hole close to extremality possesses in general an AdS$_2$ factor. 
While the mass of the scalar field can be above
the BF bound associated with AdS$_5$ asymptotically,
it can be below the BF bound
associated with the AdS$_2$ of the horizon.
Thus the horizon can become unstable with respect to
scalar hair formation.

Finally, Betti Hartmann considered charged Gauss-Bonnet black holes
with spherical horizon topology
\cite{Brihaye:2012cb}.
Again an instability related to the condensation of a scalar field was observed.
Besides it was seen that hairy Gauss-Bonnet black holes 
never tend to regular soliton solutions when the horizon radius
tends to zero.

\subsection{New solution generation techniques using SL(2,R)-duality
and general black holes}

To obtain new analytical black hole solutions various
solution generation techniques are known.
A straightforward method to obtain new black hole
solutions in $D$ dimensions is based on the Kaluza-Klein reduction.
Here one embeds
a $D$-dimensional vacuum solution in $D+1$ dimensions, 
%performs a boost with respect to time and the additional coordinate, 
performs a symmetry transformation,
and then reduces the solution to $D$ dimensions
\cite{Maison:1979kx,Chodos:1980df,Dobiasch:1981vh,Gibbons:1985ac,Gibbons:1987ps,Frolov:1987rj,Breitenlohner:1987dg,Horne:1992zy,Rasheed:1995zv,Maison:2000fj,Kunz:2006jd}.
In the simplest case, when a single boost is applied, the procedure leads to electrically
charged Einstein-Maxwell-dilaton black holes in $D$ dimensions, obtained
in four dimensions by Chodos and Detweiler \cite{Chodos:1980df} in the static case,
and by Frolov et al.~\cite{Frolov:1987rj} in the rotating case.
 
By applying two boosts and a rotation Rasheed \cite{Rasheed:1995zv} obtained
the most general electrically and magnetically charged rotating black hole solutions 
of Einstein-Maxwell-dilaton theory in four dimensions.
From a five-dimensional perspective, however, it corresponds to a solution
of pure Einstein gravity.
This solution plays an important role in the construction of new solutions
performed by Shinya Tomizawa, as he explained in his talk.

A shown in \cite{Chamseddine:1980sp}
the Kaluza-Klein reduction of five-dimensional minimal supergravity to four
dimensions yields a theory with a dilaton, an axion, and two gauge fields.
The SL(2, $R$) symmetry of the resulting set of equations of motion
was utilized by Shinya Tomizawa \cite{Mizoguchi:2011zj,Mizoguchi:2012vg,Tomizawa:2012nk}
  %``New approach to solution generation using SL(2,R)-duality of a dimensionally reduced space in five-dimensional minimal supergravity and new black holes,''
to generate new solutions in five-dimensional minimal supergravity.
From the four-dimensional point of view
these black hole solutions carry six different global charges. These are the
mass, the angular momentum, the Kaluza-Klein electric and magnetic charge, and 
the electric and magnetic charge of the Maxwell field.

The previously known black hole solutions of this theory
were static \cite{Elvang:2005sa,Gaiotto:2005xt,Ishihara:2005dp,Nakagawa:2008rm,Tomizawa:2008hw}
with one exception \cite{Tomizawa:2008qr}.
Moreover, none of the known solutions did carry all six charges.
With the new solution generation technique presented in this talk
finally black hole solutions with all six charges could be generated
\cite{Mizoguchi:2011zj,Mizoguchi:2012vg,Tomizawa:2012nk}.

First the Rasheed solution \cite{Rasheed:1995zv} represented the {\sl seed} solution,
i.e., the known solution, on which a number of SL(2, $R$) transformations
were performed to obtain new solutions \cite{Mizoguchi:2011zj}.
However, in these new solutions four of the six global charges,
the Kaluza-Klein electric and magnetic charge as well as the electric and
magnetic charge of the Maxwell field, are connected by a
relation. Thus the six global charges of these solutions are not fully
independent. 

Subsequently,
another seed solution was employed,
that corresponds to a boosted rotating electrically and magnetically
charged black string solution with five parameters
\cite{Compere:2010fm}.
This seed then led to the most general set of black hole
solutions of the theory \cite{Tomizawa:2012nk}.
These are specified
by six {\sl independent} global charges.

\subsection{New squashed black hole solution with electromagnetic field}

Another interesting type of solution with compact dimensions
are the Kaluza-Klein black hole solutions
with squashed horizons,
where the squashing transformation may also be viewed as
a solution generation technique.
Squashed black holes in five dimensions were first constructed
by Ishihara and Matsuno \cite{Ishihara:2005dp}.
Possessing an $S^3$ horizon topology,
these black hole solutions asymptote to a locally flat space-time,
which is a twisted $S^1$ fiber bundle over the four-dimensional
Minkowski space-time.
See \cite{Tomizawa:2011mc} for an overview of 
black hole solutions with squashed horizons.

In his talk Stoytcho Yazadjiev first presented  a new squashed
Kaluza-Klein black hole solution.
It represents a solution of Einstein-Maxwell-dilation theory
in five dimensions, obtained from a Kaluza-Klein reduction
of six-dimensional Einstein gravity.
Its various limits represent the 
Ishihara-Matsuno solution \cite{Ishihara:2005dp},% in the vaccum case,
and Einstein-Maxwell-dilation generalizations
of the dipole black string and 
the Gross-Perry-Sorkin
Kaluza-Klein monopole \cite{Gross:1983hb,Sorkin:1983ns}.
According to the topology theorem \cite{Hollands:2007aj,Hollands:2007qf}
the horizon topology of this solution is determined by 
its interval structure.
The uniqueness theorem for Einstein-Maxwell-dilation theory
with compact extra dimensions 
\cite{Hollands:2008fm,Yazadjiev:2010uu} furthermore states, 
that this solution is fully characterized by its interval structure 
%(with respect to canonical basis!) 
and by its magnetic fluxes.

With the presentation of his second new solution 
Stoytcho Yazadjiev turned 
to Kaluza-Klein multi-black hole configurations
(see e.g.~the review \cite{Tomizawa:2011mc}).
In particular, he presented 
a solution describing
black lenses in equilibrium in Einstein-Maxwell-dilation theory
in five dimensions
\cite{Yazadjiev:2012yu}.
  %``Kaluza-Klein rotating multi-black hole configurations with electromagnetic field in Einstein-Maxwell-dilaton gravity,''

Interestingly, an old solution by Clement \cite{Clement:1985gm}
had been reconsidered recently by Matsuno et al.~\cite{Matsuno:2012hf}
and interpreted as an equilibrium configuration of extremal rotating
black holes, held apart by their repulsive spin-spin interaction.
Subsequently, they generalized
this solution to five-dimensional Einstein-Maxwell theory
and minimal supergravity \cite{Matsuno:2012ge}.
The generalization by Stoytcho Yazadjiev, on the other hand,
gave a solution that describes several black lenses in equilibrium,
where the electric charge of each black lense is zero.
Moreover, the horizons are superconducting 
in the sense that they expel the magnetic flux lines
\cite{Bicak:1980du,Chamblin:1998qm}.
The gravitational force between the black lenses is balanced by the
tension of the compact dimension and their repulsive spin-spin
interaction.

\subsection{Charged black holes on the Taub-bolt instanton
and their thermodynamics}

Besides multi-black hole solutions, also
sequences of Kaluza-Klein bubbles and black holes were
analyzed in recent years 
\cite{Elvang:2002br,Elvang:2004iz,Tomizawa:2007mz,Kunz:2008rs,Yazadjiev:2009nm,Yazadjiev:2009gr,Nedkova:2010gn}.
Here the bubbles provide a means to hold the black holes apart,
allowing for multi-black hole space-times without conical singularites.
At the same time, the bubbles may be viewed 
as the simplest case of gravitational instantons.

Associated with the Gross-Perry-Sorkin magnetic monopole
\cite{Gross:1983hb,Sorkin:1983ns},
the self-dual Taub-NUT instanton represents another gravitational instanton.
The squashed black holes \cite{Ishihara:2005dp}
may be viewed as black holes sitting on this instanton.
Further instanton backgrounds are Euclidean Kerr, Taub-bolt
or Eguchi-Hanson instantons.
Chen and Teo employed the rod-structure formalism 
to provide a classification of gravitational instantons
\cite{Chen:2010zu},
based on the classification theorems 
\cite{Hollands:2007qf,Hollands:2008fm}.
They also provided a recent review on
black holes on gravitational instantons \cite{Chen:2010ih}.

In her talk Petya Nedkova first recalled the partial classifications
of gravitational instantons by means of 
(i) the asymptotic space-time structure \cite{Gibbons:1979gd},
and
(ii) the fixed point sets of the U(1) subgroup 
of the isometry group \cite{Gibbons:1979xm},
as well as the complete classification by means of
the fixed point sets of the full U(1) $\times$ U(1) isometry group
\cite{Hollands:2007aj,Hollands:2007qf}.
Then she discussed the generalization obtained
by including event horizons, i.e.,~black holes `sitting' on
gravitational instantons.

Here her main focus was the analysis of the physical properties of
these solutions.
In particular, extending previous work 
\cite{Mann:2005cx,Astefanesei:2005ad,Townsend:2001rg,Traschen:2001pb,Traschen:2003jm,Kastor:2008wd}
Petya Nedkova derived the expressions for the mass, the tension,
the NUT charge and the NUT potential, and she
obtained new Smarr-like relations and the first law
for the mass and the tension
\cite{Nedkova:2011hx}.
  %``On the Thermodynamics of 5D Black Holes on ALF Gravitational Instantons,''
The culmination was the presentation of a new solution,
obtained with the solution generation technique \cite{Yazadjiev:2005hr}.
The seed solution is a static black hole on a Taub-bolt instanton
\cite{Chen:2010ih},
which is transformed into a charged black hole on the Taub-bolt instanton
of Einstein-Maxwell-dilaton theory
\cite{Nedkova:2011hx}.
The new thermodynamical formulae then yielded the
Smarr relations and the first law for this solution.

\subsection{Black rings on Taub-NUT}

The Emparan-Reall \cite{Emparan:2001wn} 
and the Pomeransky-Sen'kov \cite{Pomeransky:2006bd}
black rings are singly and doubly rotating 
asymptotically flat vacuum solutions.
But like black holes, also black rings may be
considered on other backgrounds (see e.g.~\cite{Chen:2010ih}).
Black rings on bubbles, for instance,
were considered by Yazadjiev and Nedkova \cite{Yazadjiev:2009gr,Nedkova:2010gn}.
In his talk Yu Chen considered
vacuum black rings on Taub-NUT backgrounds
\cite{Chen:2012zb}.
  %``Rotating black rings on Taub-NUT,''

In contrast to black rings on Taub-NUT backgrounds 
in supersymmetric theories \cite{Elvang:2005sa,Gaiotto:2005xt,Bena:2005ni},
vacuum black rings on Taub-NUT backgrounds
proved to be more difficult to construct.
While the static solutions of Ford et al.~\cite{Ford:2007th}
exhibited conical singularities,
in the rotating solutions of Camps \cite{Camps:2008hb}
balance could be achieved. However, the
rotations with respect to the $S^1$ and the $S^2$
directions of the ring were not independent.
It was the goal of Chen and Teo to obtain 
the most general balanced rotating black rings on Taub-NUT
\cite{Chen:2012zb}.

As explained by Yu Chen in his talk,
their construction of black rings on Taub-NUT
was based on the inverse scattering method 
\cite{Belinski,Pomeransky:2005sj}.
He first discussed the singly rotating black rings on Taub-NUT.
These possess the correct rod structure and
may be viewed as the direct generalizations of the
Emparan-Reall black rings, which are recovered
in the limit of infinite NUT charge.
When balance is imposed, the solutions are
completely regular: they possess no naked singularities,
no closed timelike curves, no Dirac-Misner string
\cite{Chen:2012zb}.

It is very interesting to perform a Kaluza-Klein reduction
on these black rings along the direction of the
Killing vector that generates the finite circle at infinity.
Then the NUT of the space-time reduces to the 
Gross-Perry-Sorkin monopole \cite{Gross:1983hb,Sorkin:1983ns},
and the black ring reduces to an electrically charged black hole.
Thus the magnetic charge is carried by the monopole,
while the electric charge is carried by the black hole.
The angular momentum of the system then arises
only from the electromagnetic field, and is given
by the product of the electric and the magnetic charge.
In general, the monopole and the black hole are
separated by a conical singularity,
but the reduced four-dimensional system is balanced,
when balance was achieved for the five-dimensional solution
\cite{Chen:2012zb}.

Yu Chen then addressed the Taub-NUT generalizations of the
Pomeransky-Sen'kov black rings. 
In particular, he presented two classes of extremal solutions,
where the $S^2$ rotation is saturated so that
the surface gravity vanishes
\cite{Chen:2012zb}.
The sign of the $S^2$ rotation distinguishes the two classes,
which represent physically different configurations 
for finite NUT charge. But both classes reduce
to the extremal Pomeransky-Sen'kov black rings
in the infinite NUT charge limit.
A Kaluza-Klein reduction now
yielded a purely electrically charged rotating
extremal black hole, that can be kept in balance
with a magnetic monopole.
The extremal black hole is associated with the Rasheed solution
\cite{Rasheed:1995zv}.
Depending on the sign of the rotation of the $S^2$,
the black hole may be considered as corotating or counterrotating
with respect to the angular momentum carried by
the electromagnetic field.

\section{New investigations of the properties of higher-dimensional solutions}

We now turn to the discussion of physical and mathematical 
properties of higher-dimensional black hole solutions,
starting with a new approach to address their stability.

\subsection{Black hole instabilities and local Penrose inequalities}

While a stability analysis of higher dimensional black holes
is very important from a pure gravity point of view, 
the gravity/gauge duality presents a further motivation.
As discussed above, a stability analysis of AdS black holes can help
understand the phase structure of the corresponding boundary field theory.

A linear stability analysis
of the static Schwarzschild-Tangherlini black holes
showed them to be stable \cite{Kodama:2003jz,Ishibashi:2003ap,Konoplya:2007jv}.
A stability analysis for rotating higher dimensional black holes, however,
is much more involved,
since the perturbation equations are partial differential equations,
where the separation of variables presents a basic difficulty
\cite{Kunduri:2006qa,Murata:2007gv,Murata:2008yx}.
So far basically the cohomogeneity-1 case of black holes
with equal magnitude angular momenta and the singly spinning case
\cite{Kodama:2009bf,Dias:2009iu,Dias:2010eu,Dias:2010maa,Dias:2011jg,Shibata:2009ad,Shibata:2010wz}
were analyzed.
As predicted by Emparan and Myers \cite{Emparan:2003sy},
the Gregory-Laflamme type instabilities \cite{Gregory:1993vy}
were found \cite{Dias:2009iu,Dias:2010eu,Dias:2010maa,Dias:2011jg}, 
but also bar-mode instabilities were discovered
\cite{Shibata:2009ad,Shibata:2010wz}.
(See \cite{Murata:2011zz} for a recent review.)

In his talk Keiju Murata presented a new and simpler method
to investigate the stability of black holes
\cite{Figueras:2011he}.
This method allows to demonstrate %certain types of
instability by employing inequalities analogous to the Penrose inequality
(see \cite{Mars:2009cj} for a recent review).
Keiju Murata explained, that if a certain stationary black hole is stable,
then initial data describing a small perturbation of the black hole
must satisfy a local Penrose inequality, and that a black hole
must be unstable if one can find initial data that violates this
inequality 
\cite{Figueras:2011he}.

In the derivation of the Penrose inequality a Taylor expansion for
a one-parameter family of solutions is considered.
For rotating black holes with axial symmetry
the inequality then reads
$$Q = \ddot M(0) - \frac{1}{4} T \ddot A_{\rm app}(0) -
\frac{1}{T c_J} \dot M(0)^2 \ge 0 $$
where the derivative is with respect to the small parameter,
and $M$ denotes the mass, $T$ the temperature, $A_{\rm app}$
the apparent horizon area and $c_J$ the heat capacity at constant
angular momentum
\cite{Figueras:2011he}.
According to Keiju Murata the strategy is then to
construct initial data describing a black hole perturbation,
to read off the deviation of the mass and the area of apparent horizon,
to substitute them into $Q$, and to check the sign.
If $Q<0$, the black hole is unstable.

His first application were black strings, where the inequality simplifies
to $\ddot A_{\rm app}(0) > 0$.
Although this technique gives only a sufficient condition for instability,
the results are amazingly close to the exact values
for the Gregory-Laflamme instability \cite{Gregory:1993vy}.
For singly spinning Myers-Perry black holes, on the other hand,
the results are consistent with the numerical results
\cite{Dias:2009iu,Dias:2010eu,Dias:2010maa,Dias:2011jg},
but the instability is seen only at notably higher values of the spin.
When applied to black rings, the new method showed, that
singly and doubly spinning black rings are unstable on their
fat black ring branch
\cite{Figueras:2011he}.
The method should prove powerful also in many other cases,
and in particular in asymptotically AdS space-times.

\subsection{Existence and some properties of
thermodynamic black di-rings}

We now address some properties of multi-black objects.
Balanced black Saturns,
for instance, are regular stationary solutions \cite{Elvang:2007rd}.
In general, these two-component solutions
exhibit a two-fold continuous non-uniqueness.
However, by imposing thermodynamic equilibrium on the system,
i.e., by requiring that the black hole and the black ring 
have equal temperature and equal angular velocity,
one drastically reduces the phase space \cite{Elvang:2007hg}.
In particular, this selects a single family of solutions,
consisting of a thin black Saturn branch
and a fat black Saturn branch \cite{Elvang:2007hg}.

Another such composite system are black di-rings.
In his poster presentation
Takashi Mishima recalled that black di-rings
were obtained with two different methods:
first, by employing a solution generating technique
similar to a B\"acklund transformation
\cite{Iguchi:2007is},
and second, by the inverse scattering method,
employing two different seeds
\cite{Evslin:2007fv,Iguchi:2011qi}.

After clarifying that the black di-rings obtained
by the two methods are completely equivalent
\cite{Iguchi:2010pe},
Takashi Mishima addressed the question,
whether black di-rings can exist in thermodynamic equilibrium.
Indeed, as for the black Saturn,
there exists a single family of solutions,
consisting of a thin black di-ring branch
and a fat black di-ring branch
\cite{Iguchi:2010pe}.
  %``Thermodynamic black di-rings,''
The fat black di-ring branch ends in a singular configuration
at the same point, where the Myers-Perry and black Saturn branches end.
The thin black di-ring branch, on the other hand,
has a limit, where the ratio of the total area 
of the black di-ring to that of the black ring
becomes one-half.

\subsection{Analyticity of event horizons of extremal Kaluza-Klein
black holes}

As argued by Masashi Kimura in his talk, 
analyticity of the metric is a crucial property for the investigation of the 
global structure of the space-time
\cite{Kimura:2008cq}.
  %``Analyticity of Event Horizons of Five-Dimensional Multi-Black Holes with Non-Trivial Asymptotic Structure,''
In particular, he addressed the analyticity of the horizon for extremal
Kaluza-Klein multi-black holes.

In four dimensions the Majumdar-Papapetrou solution \cite{Majumdar:1947eu,Papaetrou:1947ib} 
describes a set of extremal charged static black holes, where
the gravitational and the Coulomb forces exactly balance.
These solutions were extended to multi-black holes in higher dimensions by Myers
\cite{Myers:1986rx}.
Whereas the black holes of the Majumdar-Papapetrou solution 
possess analytic event horizons \cite{Hartle:1972ya},
the event horizons of the higher-dimensional multi-black holes \cite{Myers:1986rx}
are not analytic \cite{Gibbons:1994vm,Welch:1995dh,Candlish:2007fh}.

Masashi Kimura then considered a toy model for extremal Kaluza-Klein multi-black holes.
In particular, he suggested to superpose the black holes periodically 
and to make identifications,
either with respect to the single period $l$ in the case of $S^1$, 
or with respect to all $N$ periods $l^{(n)}$ in the case of $T^N$  extra dimensions.
He explained that
(i) for two black holes
the metric is $C^2$, but not $C^3$ at the horizon in five dimensions,
while it is $C^1$, but not $C^2$ in more than five dimensions;
(ii) for $S^1$ and $T^N$ compactifications
the metric is analytic in five dimensions,
while it is again $C^1$, but not $C^2$ in more than five dimensions.

\subsection{Chaos in geodesic motion around a black ring}

To understand the physical properties of black hole solutions it is essential to study the orbits
of test particles and light in these space-times.
The geodesic equations for test particles and light 
in Myers-Perry black hole space-times are separable 
\cite{Kubiznak:2006kt,Page:2006ka,Frolov:2006pe}. 
Geodesics in Myers-Perry black hole space-times were studied in
\cite{Frolov:2003en,Gooding:2008tf,Hackmann:2008tu,Kagramanova:2012hw}.

In his talk Takahisa Igata addressed the geodesics of black ring space-times,
where separability does not appear to hold in general.
Employing ring coordinates,
the equations of motion could be separated only in special cases.
These correspond to either
geodesics on the two rotational axes, or to zero energy null geodesics, 
which can only exist in the ergosphere \cite{Hoskisson:2007zk,Durkee:2008an}.
In these special cases 
the geodesic motion was studied numerically and analytically
in the singly spinning black ring space-times 
 \cite{Hoskisson:2007zk,Elvang:2006dd,Igata:2010ye,Armas:2010pw,Grunau:2012ai},
as well as in the doubly spinning  black ring space-times 
\cite{Durkee:2008an,Grunau:2012ri}.
Interestingly, in contrast to Myers-Perry black hole space-times,
there are stable bound orbits in black ring space-times
\cite{Igata:2010ye,Grunau:2012ai,Grunau:2012ri,Igata:2013be}.

Separability of the Hamilton-Jacobi equation depends on the coordinate system.
Therefore one might try to think of a coordinate system 
in which it would be possible to separate the Hamilton-Jacobi equation 
for general geodesics. 
To settle the issue, Takahisa Igata analyzed 
the geometry of the singly spinning black rings
further in his talk. By using the Poincar\'e  map
he argued that these ring space-times possess chaotic bound orbits 
\cite{Igata:2010cd}.
  %``Chaos in Geodesic Motion around a Black Ring,''
The appearance of chaos, however, implies that there is no
additional constant of motion in the black ring metric.
Thus it should not be possible to separate the Hamilton-Jacobi equation 
of singly spinning black rings in any coordinate system, in general.
It will be interesting to see whether chaotic motion will also appear in the
doubly spinning black ring space-times.

\section{Conclusion}

The field is still very active and attracting many excellent young
researchers, as seen in this and related sessions.
Clearly, higher-dimensional black objects are still in the focus
of the attention of the community.


\begin{thebibliography}{1000}
\itemsep=0pt

%\cite{Myers:1986un}
\bibitem{Myers:1986un}
  R.~C.~Myers and M.~J.~Perry,
  %``Black Holes In Higher Dimensional Space-Times,''
  Annals Phys.\  {\bf 172}, 304 (1986).
  %%CITATION = APNYA,172,304;%%

%\cite{Emparan:2001wn}
\bibitem{Emparan:2001wn}
  R.~Emparan and H.~S.~Reall,
  %``A rotating black ring in five dimensions,''
  Phys.\ Rev.\ Lett.\  {\bf 88}, 101101 (2002)
  [arXiv:hep-th/0110260].
  %%CITATION = PRLTA,88,101101;%%

%\cite{Emparan:2008eg}
\bibitem{Emparan:2008eg}
  R.~Emparan and H.~S.~Reall,
  %``Black Holes in Higher Dimensions,''
  Living Rev.\ Rel.\  {\bf 11}, 6 (2008)
  [arXiv:0801.3471 [hep-th]].
  %%CITATION = 00222,11,6;%%




%\cite{Emparan:2007wm}
\bibitem{Emparan:2007wm}
  R.~Emparan, T.~Harmark, V.~Niarchos, N.~A.~Obers and M.~J.~Rodriguez,
  %``The Phase Structure of Higher-Dimensional Black Rings and Black Holes,''
  JHEP {\bf 0710} (2007) 110
  [arXiv:0708.2181 [hep-th]].

%\cite{Emparan:2009vd}
\bibitem{Emparan:2009vd}
  R.~Emparan, T.~Harmark, V.~Niarchos and N.~A.~Obers,
  %``New Horizons for Black Holes and Branes,''
  JHEP {\bf 1004}, 046 (2010)
  [arXiv:0912.2352 [hep-th]];

  %\cite{Emparan:2009cs}
\bibitem{Emparan:2009cs}
  R.~Emparan, T.~Harmark, V.~Niarchos and N.~A.~Obers,
  %``Blackfolds,''
  Phys.\ Rev.\ Lett.\  {\bf 102} (2009) 191301
  [arXiv:0902.0427 [hep-th]].

%\cite{Kleihaus:2012xh}
\bibitem{Kleihaus:2012xh} 
  B.~Kleihaus, J.~Kunz and E.~Radu,
  %``Black rings in six dimensions,''
  Phys.\ Lett.\ B {\bf 718}, 1073 (2013)
  [arXiv:1205.5437 [hep-th]].

%\cite{Emparan:2010sx}
\bibitem{Emparan:2010sx}
  R.~Emparan and P.~Figueras,
  %``Multi-black rings and the phase diagram of higher-dimensional black
  %holes,''
  JHEP {\bf 1011} (2010) 022
  [arXiv:1008.3243 [hep-th]].

%\cite{Gregory:1993vy}
\bibitem{Gregory:1993vy}
  R.~Gregory and R.~Laflamme,
  %``Black strings and p-branes are unstable,''
  Phys.\ Rev.\ Lett.\  {\bf 70}, 2837 (1993)
  [arXiv:hep-th/9301052]

%\cite{Emparan:2003sy}
\bibitem{Emparan:2003sy}
  R.~Emparan and R.~C.~Myers,
  %``Instability of ultra-spinning black holes,''
  JHEP {\bf 0309}, 025 (2003)
  [hep-th/0308056].

%\cite{Dias:2009iu}
\bibitem{Dias:2009iu}
  O.~J.~C.~Dias, P.~Figueras, R.~Monteiro, J.~E.~Santos and R.~Emparan,
  %``Instability and new phases of higher-dimensional rotating black holes,''
  Phys.\ Rev.\ D {\bf 80} (2009) 111701
  [arXiv:0907.2248 [hep-th]].
  %%CITATION = ARXIV:0907.2248;%%

%\cite{Dias:2010eu}
\bibitem{Dias:2010eu}
  O.~J.~C.~Dias, P.~Figueras, R.~Monteiro, H.~S.~Reall and J.~E.~Santos,
  %``An instability of higher-dimensional rotating black holes,''
  JHEP {\bf 1005}, 076 (2010)
  [arXiv:1001.4527 [hep-th]].

  %\cite{Dias:2010maa}
\bibitem{Dias:2010maa}
  O.~J.~C.~Dias, P.~Figueras, R.~Monteiro and J.~E.~Santos,
  %``Ultraspinning instability of rotating black holes,''
  Phys.\ Rev.\ D {\bf 82}, 104025 (2010)
  [arXiv:1006.1904 [hep-th]].
  %%CITATION = ARXIV:1006.1904;%% 

%\cite{Dias:2011jg}
\bibitem{Dias:2011jg} 
  O.~J.~C.~Dias, R.~Monteiro and J.~E.~Santos,
  %``Ultraspinning instability: the missing link,''
  JHEP {\bf 1108}, 139 (2011)
  [arXiv:1106.4554 [hep-th]].

%\cite{Kleihaus:2009wh}
\bibitem{Kleihaus:2009wh} 
  B.~Kleihaus, J.~Kunz and E.~Radu,
  %``d greater than or equal to five static black holes with S**2 x S**(d-4) event horizon topology,''
  Phys.\ Lett.\ B {\bf 678}, 301 (2009)
  [arXiv:0904.2723 [hep-th]].

%\cite{Kleihaus:2010pr}
\bibitem{Kleihaus:2010pr} 
  B.~Kleihaus, J.~Kunz, E.~Radu and M.~J.~Rodriguez,
  %``New generalized nonspherical black hole solutions,''
  JHEP {\bf 1102}, 058 (2011)
  [arXiv:1010.2898 [gr-qc]].

%\cite{Kleihaus:2013zpa}
\bibitem{Kleihaus:2013zpa} 
  B.~Kleihaus, J.~Kunz and E.~Radu,
  %``$d\geq 5$ magnetized static, balanced black holes with $S^2\times S^{d-4}$ event horizon topology,''
  Phys.\ Lett.\ B {\bf 723}, 182 (2013)
  [arXiv:1303.2190 [gr-qc]].


%\cite{Maldacena:1997re}
\bibitem{Maldacena:1997re} 
  J.~M.~Maldacena,
  %``The Large N limit of superconformal field theories and supergravity,''
  Adv.\ Theor.\ Math.\ Phys.\  {\bf 2}, 231 (1998)
  [hep-th/9711200].

%\cite{Gubser:2008px}
\bibitem{Gubser:2008px} 
  S.~S.~Gubser,
  %``Breaking an Abelian gauge symmetry near a black hole horizon,''
  Phys.\ Rev.\ D {\bf 78}, 065034 (2008)
  [arXiv:0801.2977 [hep-th]].

%\cite{Hartnoll:2008vx}
\bibitem{Hartnoll:2008vx} 
  S.~A.~Hartnoll, C.~P.~Herzog and G.~T.~Horowitz,
  %``Building a Holographic Superconductor,''
  Phys.\ Rev.\ Lett.\  {\bf 101}, 031601 (2008)
  [arXiv:0803.3295 [hep-th]].

%\cite{Hartnoll:2008kx}
\bibitem{Hartnoll:2008kx} 
  S.~A.~Hartnoll, C.~P.~Herzog and G.~T.~Horowitz,
  %``Holographic Superconductors,''
  JHEP {\bf 0812}, 015 (2008)
  [arXiv:0810.1563 [hep-th]].

%\cite{Horowitz:2008bn}
\bibitem{Horowitz:2008bn} 
  G.~T.~Horowitz and M.~M.~Roberts,
  %``Holographic Superconductors with Various Condensates,''
  Phys.\ Rev.\ D {\bf 78}, 126008 (2008)
  [arXiv:0810.1077 [hep-th]].

%\cite{Breitenlohner:1982bm}
\bibitem{Breitenlohner:1982bm} 
  P.~Breitenlohner and D.~Z.~Freedman,
  %``Positive Energy in anti-De Sitter Backgrounds and Gauged Extended Supergravity,''
  Phys.\ Lett.\ B {\bf 115}, 197 (1982).

%\cite{Breitenlohner:1982jf}
\bibitem{Breitenlohner:1982jf} 
  P.~Breitenlohner and D.~Z.~Freedman,
  %``Stability in Gauged Extended Supergravity,''
  Annals Phys.\  {\bf 144}, 249 (1982).

%\cite{Gregory:2009fj}
\bibitem{Gregory:2009fj} 
  R.~Gregory, S.~Kanno and J.~Soda,
  %``Holographic Superconductors with Higher Curvature Corrections,''
  JHEP {\bf 0910}, 010 (2009)
  [arXiv:0907.3203 [hep-th]].

%\cite{Brihaye:2010mr}
\bibitem{Brihaye:2010mr} 
  Y.~Brihaye and B.~Hartmann,
  %``Holographic Superconductors in 3+1 dimensions away from the probe limit,''
  Phys.\ Rev.\ D {\bf 81}, 126008 (2010)
  [arXiv:1003.5130 [hep-th]].

%\cite{Brihaye:2011hm}
\bibitem{Brihaye:2011hm} 
  Y.~Brihaye and B.~Hartmann,
  %``Stability of Gauss-Bonnet black holes in Anti-de-Sitter space-time against scalar field condensation,''
  Phys.\ Rev.\ D {\bf 84}, 084008 (2011)
  [arXiv:1107.3384 [gr-qc]].

%\cite{Brihaye:2012cb}
\bibitem{Brihaye:2012cb} 
  Y.~Brihaye and B.~Hartmann,
  %``Hairy charged Gauss-Bonnet solitons and black holes,''
  Phys.\ Rev.\ D {\bf 85}, 124024 (2012)
  [arXiv:1203.3109 [gr-qc]].


%\cite{Maison:1979kx}
\bibitem{Maison:1979kx} 
  D.~Maison,
  %``Ehlers-harrison Type Transformations For Jordan's Extended Theory Of Gravitation,''
  Gen.\ Rel.\ Grav.\  {\bf 10}, 717 (1979).

%\cite{Maison:2000fj}
\bibitem{Maison:2000fj} 
  D.~Maison,
  %``Duality and hidden symmetries in gravitational theories,''
  Lect.\ Notes Phys.\  {\bf 540}, 273 (2000).

%\cite{Chodos:1980df}
\bibitem{Chodos:1980df} 
  A.~Chodos and S.~L.~Detweiler,
  %``Spherically Symmetric Solutions in Five-dimensional General Relativity,''
  Gen.\ Rel.\ Grav.\  {\bf 14}, 879 (1982).

%\cite{Dobiasch:1981vh}
\bibitem{Dobiasch:1981vh} 
  P.~Dobiasch and D.~Maison,
  %``Stationary, Spherically Symmetric Solutions of Jordan's Unified Theory of Gravity and Electromagnetism,''
  Gen.\ Rel.\ Grav.\  {\bf 14}, 231 (1982).

%\cite{Gibbons:1985ac}
\bibitem{Gibbons:1985ac} 
  G.~W.~Gibbons and D.~L.~Wiltshire,
  %``Black Holes in Kaluza-Klein Theory,''
  Annals Phys.\  {\bf 167}, 201 (1986)
  [Erratum-ibid.\  {\bf 176}, 393 (1987)].

%\cite{Gibbons:1987ps}
\bibitem{Gibbons:1987ps} 
  G.~W.~Gibbons and K.~-i.~Maeda,
  %``Black Holes and Membranes in Higher Dimensional Theories with Dilaton Fields,''
  Nucl.\ Phys.\ B {\bf 298}, 741 (1988).

%\cite{Frolov:1987rj}
\bibitem{Frolov:1987rj} 
  V.~P.~Frolov, A.~I.~Zelnikov and U.~Bleyer,
  %``Charged Rotating Black Hole From Five-dimensional Point of View,''
  Annalen Phys.\  {\bf 44}, 371 (1987).

%\cite{Breitenlohner:1987dg}
\bibitem{Breitenlohner:1987dg}
  P.~Breitenlohner, D.~Maison and G.~W.~Gibbons,
  %``Four-Dimensional Black Holes from Kaluza-Klein Theories,''
  Commun.\ Math.\ Phys.\  {\bf 120}, 295 (1988).


%\cite{Horne:1992zy}
\bibitem{Horne:1992zy} 
  J.~H.~Horne and G.~T.~Horowitz,
  %``Rotating dilaton black holes,''
  Phys.\ Rev.\ D {\bf 46}, 1340 (1992)
  [hep-th/9203083].

%\cite{Rasheed:1995zv}
\bibitem{Rasheed:1995zv} 
  D.~Rasheed,
  %``The Rotating dyonic black holes of Kaluza-Klein theory,''
  Nucl.\ Phys.\ B {\bf 454}, 379 (1995)
  [hep-th/9505038].

%\cite{Kunz:2006jd}
\bibitem{Kunz:2006jd} 
  J.~Kunz, D.~Maison, F.~Navarro-Lerida and J.~Viebahn,
  %``Rotating Einstein-Maxwell-dilaton black holes in D dimensions,''
  Phys.\ Lett.\ B {\bf 639}, 95 (2006)
  [hep-th/0606005].

%\cite{Chamseddine:1980sp}
\bibitem{Chamseddine:1980sp} 
  A.~H.~Chamseddine and H.~Nicolai,
  %``Coupling the SO(2) Supergravity Through Dimensional Reduction,''
  Phys.\ Lett.\ B {\bf 96}, 89 (1980).

%\cite{Mizoguchi:2011zj}
\bibitem{Mizoguchi:2011zj} 
  S.~Mizoguchi and S.~Tomizawa,
  %``New approach to solution generation using SL(2,R)-duality of a dimensionally reduced space in five-dimensional minimal supergravity and new black holes,''
  Phys.\ Rev.\ D {\bf 84}, 104009 (2011)
  [arXiv:1106.3165 [hep-th]].

%\cite{Mizoguchi:2012vg}
\bibitem{Mizoguchi:2012vg} 
  S.~Mizoguchi and S.~Tomizawa,
  %``Flipped $SL(2,R)$ duality in five-dimensional supergravity,''
  Phys.\ Rev.\ D {\bf 86}, 024022 (2012)
  [arXiv:1201.3063 [hep-th]].

%\cite{Tomizawa:2012nk}
\bibitem{Tomizawa:2012nk} 
  S.~Tomizawa and S.~Mizoguchi,
  %``General Kaluza-Klein black holes with all six independent charges in five-dimensional minimal supergravity,''
  Phys.\ Rev.\ D {\bf 87}, 024027 (2013)
  [arXiv:1210.6723 [hep-th]].

%\cite{Elvang:2005sa}
\bibitem{Elvang:2005sa} 
  H.~Elvang, R.~Emparan, D.~Mateos and H.~S.~Reall,
  %``Supersymmetric 4-D rotating black holes from 5-D black rings,''
  JHEP {\bf 0508}, 042 (2005)
  [hep-th/0504125].

%\cite{Gaiotto:2005xt}
\bibitem{Gaiotto:2005xt} 
  D.~Gaiotto, A.~Strominger and X.~Yin,
  %``5D black rings and 4D black holes,''
  JHEP {\bf 0602}, 023 (2006)
  [hep-th/0504126].

%\cite{Ishihara:2005dp}
\bibitem{Ishihara:2005dp} 
  H.~Ishihara and K.~Matsuno,
  %``Kaluza-Klein black holes with squashed horizons,''
  Prog.\ Theor.\ Phys.\  {\bf 116}, 417 (2006)
  [hep-th/0510094].

%\cite{Nakagawa:2008rm}
\bibitem{Nakagawa:2008rm} 
  T.~Nakagawa, H.~Ishihara, K.~Matsuno and S.~Tomizawa,
  %``Charged Rotating Kaluza-Klein Black Holes in Five Dimensions,''
  Phys.\ Rev.\ D {\bf 77}, 044040 (2008)
  [arXiv:0801.0164 [hep-th]].

%\cite{Tomizawa:2008hw}
\bibitem{Tomizawa:2008hw} 
  S.~Tomizawa, H.~Ishihara, K.~Matsuno and T.~Nakagawa,
  %``Squashed Kerr-Godel Black Holes: Kaluza-Klein Black Holes with Rotations of Black Hole and Universe,''
  Prog.\ Theor.\ Phys.\  {\bf 121}, 823 (2009)
  [arXiv:0803.3873 [hep-th]].

%\cite{Tomizawa:2008qr}
\bibitem{Tomizawa:2008qr} 
  S.~Tomizawa, Y.~Yasui and Y.~Morisawa,
  %``Charged Rotating Kaluza-Klein Black Holes Generated by G2(2) Transformation,''
  Class.\ Quant.\ Grav.\  {\bf 26}, 145006 (2009)
  [arXiv:0809.2001 [hep-th]].

%\cite{Compere:2010fm}
\bibitem{Compere:2010fm} 
  G.~Compere, S.~de Buyl, S.~Stotyn and A.~Virmani,
  %``A General Black String and its Microscopics,''
  JHEP {\bf 1011}, 133 (2010)
  [arXiv:1006.5464 [hep-th]].






%\cite{Tomizawa:2011mc}
\bibitem{Tomizawa:2011mc} 
  S.~Tomizawa and H.~Ishihara,
  %``Exact solutions of higher dimensional black holes,''
  Prog.\ Theor.\ Phys.\ Suppl.\  {\bf 189}, 7 (2011)
  [arXiv:1104.1468 [hep-th]].

%\cite{Gross:1983hb}
\bibitem{Gross:1983hb} 
  D.~J.~Gross and M.~J.~Perry,
  %``Magnetic Monopoles in Kaluza-Klein Theories,''
  Nucl.\ Phys.\ B {\bf 226}, 29 (1983).

%\cite{Sorkin:1983ns}
\bibitem{Sorkin:1983ns} 
  R.~d.~Sorkin,
  %``Kaluza-Klein Monopole,''
  Phys.\ Rev.\ Lett.\  {\bf 51}, 87 (1983).

%\cite{Hollands:2007aj}
\bibitem{Hollands:2007aj} 
  S.~Hollands and S.~Yazadjiev,
  %``Uniqueness theorem for 5-dimensional black holes with two axial Killing fields,''
  Commun.\ Math.\ Phys.\  {\bf 283}, 749 (2008)
  [arXiv:0707.2775 [gr-qc]].

%\cite{Hollands:2007qf}
\bibitem{Hollands:2007qf} 
  S.~Hollands and S.~Yazadjiev,
  %``A Uniqueness theorem for 5-dimensional Einstein-Maxwell black holes,''
  Class.\ Quant.\ Grav.\  {\bf 25}, 095010 (2008)
  [arXiv:0711.1722 [gr-qc]].

%\cite{Hollands:2008fm}
\bibitem{Hollands:2008fm}
  S.~Hollands and S.~Yazadjiev,
  %``A Uniqueness theorem for stationary Kaluza-Klein black holes,''
  Commun.\ Math.\ Phys.\  {\bf 302}, 631 (2011)
  [arXiv:0812.3036 [gr-qc]].

%\cite{Yazadjiev:2010uu}
\bibitem{Yazadjiev:2010uu} 
  S.~S.~Yazadjiev,
  %``A Uniqueness theorem for black holes with Kaluza-Klein asymptotic in 5D Einstein-Maxwell gravity,''
  Phys.\ Rev.\ D {\bf 82}, 024015 (2010)
  [arXiv:1002.3954 [hep-th]].

%\cite{Yazadjiev:2012yu}
\bibitem{Yazadjiev:2012yu} 
  S.~S.~Yazadjiev,
  %``Kaluza-Klein rotating multi-black hole configurations with electromagnetic field in Einstein-Maxwell-dilaton gravity,''
  Phys.\ Rev.\ D {\bf 86}, 107504 (2012)
  [arXiv:1209.3488 [gr-qc]].

%\cite{Clement:1985gm}
\bibitem{Clement:1985gm}
  G.~Clement,
  %``Solutions of Five-dimensional General Relativity Without Spatial Symmetry,''
  Gen.\ Rel.\ Grav.\  {\bf 18}, 861 (1986).

%\cite{Matsuno:2012hf}
\bibitem{Matsuno:2012hf} 
  K.~Matsuno, H.~Ishihara, M.~Kimura and T.~Tatsuoka,
  %``Kaluza-Klein vacuum multi-black holes in five-dimensions,''
  Phys.\ Rev.\ D {\bf 86}, 044036 (2012)
  [arXiv:1206.4818 [hep-th]].

%\cite{Matsuno:2012ge}
\bibitem{Matsuno:2012ge} 
  K.~Matsuno, H.~Ishihara, M.~Kimura and T.~Tatsuoka,
  %``Charged rotating Kaluza-Klein multi-black holes and multi-black strings in five-dimensional Einstein-Maxwell theory,''
  Phys.\ Rev.\ D {\bf 86}, 104054 (2012)
  [arXiv:1208.5536 [hep-th]].

%\cite{Bicak:1980du}
\bibitem{Bicak:1980du} 
  J.~Bicak and L.~Dvorak,
  %``Stationary Electromagnetic Fields Around Black Holes. 3. General Solutions And The Fields Of Current Loops Near The Reissner-nordstrom Black Hole,''
  Phys.\ Rev.\ D {\bf 22}, 2933 (1980).

%\cite{Chamblin:1998qm}
\bibitem{Chamblin:1998qm} 
  A.~Chamblin, R.~Emparan and G.~W.~Gibbons,
  %``Superconducting p-branes and extremal black holes,''
  Phys.\ Rev.\ D {\bf 58}, 084009 (1998)
  [hep-th/9806017].





%\cite{Elvang:2002br}
\bibitem{Elvang:2002br}
  H.~Elvang and G.~T.~Horowitz,
  %``When black holes meet Kaluza-Klein bubbles,''
  Phys.\ Rev.\  D {\bf 67} (2003) 044015
  [arXiv:hep-th/0210303].
  %%CITATION = PHRVA,D67,044015;%%

%\cite{Elvang:2004iz}
\bibitem{Elvang:2004iz}
  H.~Elvang, T.~Harmark and N.~A.~Obers,
  %``Sequences of bubbles and holes: New phases of Kaluza-Klein black holes,''
  JHEP {\bf 0501} (2005) 003
  [arXiv:hep-th/0407050].

%\cite{Tomizawa:2007mz}
\bibitem{Tomizawa:2007mz} 
  S.~Tomizawa, H.~Iguchi and T.~Mishima,
  %``Rotating Black Holes on Kaluza-Klein Bubbles,''
  Phys.\ Rev.\ D {\bf 78}, 084001 (2008)
  [hep-th/0702207 [HEP-TH]].

%\cite{Kunz:2008rs}
\bibitem{Kunz:2008rs} 
  J.~Kunz and S.~Yazadjiev,
  %``Charged black holes on a Kaluza-Klein bubble,''
  Phys.\ Rev.\ D {\bf 79}, 024010 (2009)
  [arXiv:0811.0730 [hep-th]].

%\cite{Yazadjiev:2009nm}
\bibitem{Yazadjiev:2009nm} 
  S.~S.~Yazadjiev and P.~G.~Nedkova,
  %``Magnetized configurations with black holes and Kaluza-Klein bubbles: Smarr-like relations and first law,''
  Phys.\ Rev.\ D {\bf 80}, 024005 (2009)
  [arXiv:0904.3605 [hep-th]].

%\cite{Yazadjiev:2009gr}
\bibitem{Yazadjiev:2009gr}
  S.~S.~Yazadjiev and P.~G.~Nedkova,
  %``Sequences of dipole black rings and Kaluza-Klein bubbles,''
  JHEP {\bf 1001}, 048 (2010)
  [arXiv:0910.0938 [hep-th]].

%\cite{Nedkova:2010gn}
\bibitem{Nedkova:2010gn}
  P.~G.~Nedkova and S.~S.~Yazadjiev,
  %``Rotating black ring on Kaluza-Klein bubbles,''
  Phys.\ Rev.\ D {\bf 82}, 044010 (2010)
  [arXiv:1005.5051 [hep-th]].

%\cite{Chen:2010zu}
\bibitem{Chen:2010zu}
  Y.~Chen and E.~Teo,
  %``Rod-structure classification of gravitational instantons with U(1)xU(1) isometry,''
  Nucl.\ Phys.\ B {\bf 838}, 207 (2010)
  [arXiv:1004.2750 [gr-qc]].

%\cite{Chen:2010ih}
\bibitem{Chen:2010ih}
  Y.~Chen and E.~Teo,
  %``Black holes on gravitational instantons,''
  Nucl.\ Phys.\ B {\bf 850}, 253 (2011)
  [arXiv:1011.6464 [hep-th]].

%\cite{Gibbons:1979gd}
\bibitem{Gibbons:1979gd} 
  G.~W.~Gibbons, C.~N.~Pope and H.~Romer,
  %``Index Theorem Boundary Terms for Gravitational Instantons,''
  Nucl.\ Phys.\ B {\bf 157}, 377 (1979).

%\cite{Gibbons:1979xm}
\bibitem{Gibbons:1979xm} 
  G.~W.~Gibbons and S.~W.~Hawking,
  %``Classification of Gravitational Instanton Symmetries,''
  Commun.\ Math.\ Phys.\  {\bf 66}, 291 (1979).

%\cite{Mann:2005cx}
\bibitem{Mann:2005cx} 
  R.~B.~Mann and C.~Stelea,
  %``On the gravitational energy of the Kaluza Klein monopole,''
  Phys.\ Lett.\ B {\bf 634}, 531 (2006)
  [hep-th/0511180].

%\cite{Astefanesei:2005ad}
\bibitem{Astefanesei:2005ad} 
  D.~Astefanesei and E.~Radu,
  %``Quasilocal formalism and black ring thermodynamics,''
  Phys.\ Rev.\ D {\bf 73}, 044014 (2006)
  [hep-th/0509144].

%\cite{Townsend:2001rg}
\bibitem{Townsend:2001rg} 
  P.~K.~Townsend and M.~Zamaklar,
  %``The First law of black brane mechanics,''
  Class.\ Quant.\ Grav.\  {\bf 18}, 5269 (2001)
  [hep-th/0107228].

%\cite{Traschen:2001pb}
\bibitem{Traschen:2001pb} 
  J.~H.~Traschen and D.~Fox,
  %``Tension perturbations of black brane space-times,''
  Class.\ Quant.\ Grav.\  {\bf 21}, 289 (2004)
  [gr-qc/0103106].

%\cite{Traschen:2003jm}
\bibitem{Traschen:2003jm} 
  J.~H.~Traschen,
  %``A Positivity theorem for gravitational tension in brane space-times,''
  Class.\ Quant.\ Grav.\  {\bf 21}, 1343 (2004)
  [hep-th/0308173].

%\cite{Kastor:2008wd}
\bibitem{Kastor:2008wd} 
  D.~Kastor, S.~Ray and J.~Traschen,
  %``The Thermodynamics of Kaluza-Klein Black Hole/Bubble Chains,''
  Class.\ Quant.\ Grav.\  {\bf 25}, 125004 (2008)
  [arXiv:0803.2019 [hep-th]].

%\cite{Nedkova:2011hx}
\bibitem{Nedkova:2011hx} 
  P.~G.~Nedkova and S.~S.~Yazadjiev,
  %``On the Thermodynamics of 5D Black Holes on ALF Gravitational Instantons,''
  Phys.\ Rev.\ D {\bf 84}, 124040 (2011)
  [arXiv:1109.2838 [hep-th]].

%\cite{Yazadjiev:2005hr}
\bibitem{Yazadjiev:2005hr} 
  S.~S.~Yazadjiev,
  %``Asymptotically and non-asymptotically flat static black rings in charged dilaton gravity,''
  hep-th/0507097.

%%\cite{Nedkova:2011aa}
%\bibitem{Nedkova:2011aa} 
%  P.~G.~Nedkova and S.~S.~Yazadjiev,
%  %``Magnetized Black Hole on Taub-Nut Instanton,''
%  Phys.\ Rev.\ D {\bf 85}, 064021 (2012)
%  [arXiv:1112.3326 [hep-th]].







%\cite{Pomeransky:2006bd}
\bibitem{Pomeransky:2006bd} 
  A.~A.~Pomeransky and R.~A.~Sen'kov,
  %``Black ring with two angular momenta,''
  hep-th/0612005.

%\cite{Chen:2012zb}
\bibitem{Chen:2012zb} 
  Y.~Chen and E.~Teo,
  %``Rotating black rings on Taub-NUT,''
  JHEP {\bf 1206}, 068 (2012)
  [arXiv:1204.3116 [hep-th]].

%\cite{Bena:2005ni}
\bibitem{Bena:2005ni} 
  I.~Bena, P.~Kraus and N.~P.~Warner,
  %``Black rings in Taub-NUT,''
  Phys.\ Rev.\ D {\bf 72}, 084019 (2005)
  [hep-th/0504142].

%\cite{Ford:2007th}
\bibitem{Ford:2007th} 
  J.~Ford, S.~Giusto, A.~Peet and A.~Saxena,
  %``Reduction without reduction: Adding KK-monopoles to five dimensional stationary axisymmetric solutions,''
  Class.\ Quant.\ Grav.\  {\bf 25}, 075014 (2008)
  [arXiv:0708.3823 [hep-th]].

%\cite{Camps:2008hb}
\bibitem{Camps:2008hb} 
  J.~Camps, R.~Emparan, P.~Figueras, S.~Giusto and A.~Saxena,
  %``Black Rings in Taub-NUT and D0-D6 interactions,''
  JHEP {\bf 0902}, 021 (2009)
  [arXiv:0811.2088 [hep-th]].

%\cite{Belinski}
\bibitem{Belinski}
 V.~Belinski and E.~Verdaguer, 
 {\sl Gravitational solitons},
 Cambridge University Press, U.K (2001).

%\cite{Pomeransky:2005sj}
\bibitem{Pomeransky:2005sj} 
  A.~A.~Pomeransky,
  %``Complete integrability of higher-dimensional Einstein equations with additional symmetry, and rotating black holes,''
  Phys.\ Rev.\ D {\bf 73}, 044004 (2006)
  [hep-th/0507250].

%%\cite{Chen:2011tc}
%\bibitem{Chen:2011tc} 
%  Y.~Chen and E.~Teo,
%  %``A New AF gravitational instanton,''
%  Phys.\ Lett.\ B {\bf 703}, 359 (2011)
%  [arXiv:1107.0763 [gr-qc]].







%\cite{Kodama:2003jz}
\bibitem{Kodama:2003jz}
  H.~Kodama and A.~Ishibashi,
  %``A master equation for gravitational perturbations of maximally  symmetric
  %black holes in higher dimensions,''
  Prog.\ Theor.\ Phys.\  {\bf 110}, 701 (2003)
  [arXiv:hep-th/0305147].

%\cite{Ishibashi:2003ap}
\bibitem{Ishibashi:2003ap}
  A.~Ishibashi and H.~Kodama,
  %``Stability of higher-dimensional Schwarzschild black holes,''
  Prog.\ Theor.\ Phys.\  {\bf 110}, 901 (2003)
  [arXiv:hep-th/0305185].

%\cite{Konoplya:2007jv}
\bibitem{Konoplya:2007jv}
  R.~A.~Konoplya and A.~Zhidenko,
  %``Stability of multidimensional black holes: Complete numerical analysis,''
  Nucl.\ Phys.\  B {\bf 777}, 182 (2007)
  [arXiv:hep-th/0703231].

%\cite{Kunduri:2006qa}
\bibitem{Kunduri:2006qa}
  H.~K.~Kunduri, J.~Lucietti and H.~S.~Reall,
  %``Gravitational perturbations of higher dimensional rotating black holes:
  %Tensor Perturbations,''
  Phys.\ Rev.\  D {\bf 74}, 084021 (2006)
  [arXiv:hep-th/0606076].

%\cite{Murata:2007gv}
\bibitem{Murata:2007gv}
  K.~Murata and J.~Soda,
  %``A Note on Separability of Field Equations in Myers-Perry Spacetimes,''
  Class.\ Quant.\ Grav.\  {\bf 25}, 035006 (2008)
  [arXiv:0710.0221 [hep-th]].

%\cite{Murata:2008yx}
\bibitem{Murata:2008yx}
  K.~Murata and J.~Soda,
  %``Stability of Five-dimensional Myers-Perry Black Holes with Equal Angular
  %Momenta,''
  Prog.\ Theor.\ Phys.\  {\bf 120}, 561 (2008)
  [arXiv:0803.1371 [hep-th]].

%\cite{Kodama:2009bf}
\bibitem{Kodama:2009bf} 
  H.~Kodama, R.~A.~Konoplya and A.~Zhidenko,
  %``Gravitational stability of simply rotating Myers-Perry black holes: Tensorial perturbations,''
  Phys.\ Rev.\ D {\bf 81}, 044007 (2010)
  [arXiv:0904.2154 [gr-qc]].

%\cite{Shibata:2009ad}
\bibitem{Shibata:2009ad} 
  M.~Shibata and H.~Yoshino,
  %``Nonaxisymmetric instability of rapidly rotating black hole in five dimensions,''
  Phys.\ Rev.\ D {\bf 81}, 021501 (2010)
  [arXiv:0912.3606 [gr-qc]].

%\cite{Shibata:2010wz}
\bibitem{Shibata:2010wz} 
  M.~Shibata and H.~Yoshino,
  %``Bar-mode instability of rapidly spinning black hole in higher dimensions: Numerical simulation in general relativity,''
  Phys.\ Rev.\ D {\bf 81}, 104035 (2010)
  [arXiv:1004.4970 [gr-qc]].

%\cite{Murata:2011zz}
\bibitem{Murata:2011zz}
  K.~Murata,
  %``Chapter 7. Perturbative Stability Analysis of Higher Dimensional Rota ting Black Holes,''
  Prog.\ Theor.\ Phys.\ Suppl.\  {\bf 189}, 210 (2011).

%\cite{Figueras:2011he}
\bibitem{Figueras:2011he} 
  P.~Figueras, K.~Murata and H.~S.~Reall,
  %``Black hole instabilities and local Penrose inequalities,''
  Class.\ Quant.\ Grav.\  {\bf 28}, 225030 (2011)
  [arXiv:1107.5785 [gr-qc]].

%\cite{Mars:2009cj}
\bibitem{Mars:2009cj} 
  M.~Mars,
  %``Present status of the Penrose inequality,''
  Class.\ Quant.\ Grav.\  {\bf 26}, 193001 (2009)
  [arXiv:0906.5566 [gr-qc]].







%\cite{Elvang:2007rd}
\bibitem{Elvang:2007rd}
  H.~Elvang and P.~Figueras,
  %``Black Saturn,''
  JHEP {\bf 0705} (2007) 050
  [arXiv:hep-th/0701035].

%\cite{Elvang:2007hg}
\bibitem{Elvang:2007hg} 
  H.~Elvang, R.~Emparan and P.~Figueras,
  %``Phases of five-dimensional black holes,''
  JHEP {\bf 0705}, 056 (2007)
  [hep-th/0702111].

%\cite{Iguchi:2007is}
\bibitem{Iguchi:2007is}
  H.~Iguchi and T.~Mishima,
  %``Black di-ring and infinite nonuniqueness,''
  Phys.\ Rev.\  D {\bf 75} (2007) 064018
  [arXiv:hep-th/0701043].
  %%CITATION = PHRVA,D75,064018;%%

%\cite{Evslin:2007fv}
\bibitem{Evslin:2007fv}
  J.~Evslin and C.~Krishnan,
  %``The Black Di-Ring: An Inverse Scattering Construction,''
    Class.\ Quant.\ Grav.\  {\bf 26} (2009) 125018
  arXiv:0706.1231 [hep-th].
  %%CITATION = ARXIV:0706.1231;%%

%\cite{Iguchi:2011qi}
\bibitem{Iguchi:2011qi} 
  H.~Iguchi, K.~Izumi and T.~Mishima,
  %``Systematic solution-generation of five-dimensional black holes,''
  Prog.\ Theor.\ Phys.\ Suppl.\  {\bf 189}, 93 (2011)
  [arXiv:1106.0387 [gr-qc]].

%\cite{Iguchi:2010pe}
\bibitem{Iguchi:2010pe}
  H.~Iguchi and T.~Mishima,
  %``Thermodynamic black di-rings,''
  Phys.\ Rev.\ D {\bf 82}, 084009 (2010)
  [arXiv:1008.4290 [hep-th]].






%\cite{Kimura:2008cq}
\bibitem{Kimura:2008cq}
  M.~Kimura,
  %``Analyticity of Event Horizons of Five-Dimensional Multi-Black Holes with Non-Trivial Asymptotic Structure,''
  Phys.\ Rev.\ D {\bf 78}, 047504 (2008)
  [arXiv:0805.1125 [gr-qc]].

%\cite{Majumdar:1947eu}
\bibitem{Majumdar:1947eu} 
  S.~D.~Majumdar,
  %``A class of exact solutions of Einstein's field equations,''
  Phys.\ Rev.\  {\bf 72}, 390 (1947).

%\cite{Papaetrou:1947ib}
\bibitem{Papaetrou:1947ib} 
  A.~Papaetrou,
  %``A Static solution of the equations of the gravitational field for an arbitrary charge distribution,''
  Proc.\ Roy.\ Irish Acad.\ (Sect.\ A) A {\bf 51}, 191 (1947).

%\cite{Myers:1986rx}
\bibitem{Myers:1986rx} 
  R.~C.~Myers,
  %``Higher Dimensional Black Holes in Compactified Space-times,''
  Phys.\ Rev.\ D {\bf 35}, 455 (1987).

%\cite{Hartle:1972ya}
\bibitem{Hartle:1972ya} 
  J.~B.~Hartle and S.~W.~Hawking,
  %``Solutions of the Einstein-Maxwell equations with many black holes,''
  Commun.\ Math.\ Phys.\  {\bf 26}, 87 (1972).

%\cite{Gibbons:1994vm}
\bibitem{Gibbons:1994vm} 
  G.~W.~Gibbons, G.~T.~Horowitz and P.~K.~Townsend,
  %``Higher dimensional resolution of dilatonic black hole singularities,''
  Class.\ Quant.\ Grav.\  {\bf 12}, 297 (1995)
  [hep-th/9410073].

%\cite{Welch:1995dh}
\bibitem{Welch:1995dh} 
  D.~L.~Welch,
  %``On the smoothness of the horizons of multi - black hole solutions,''
  Phys.\ Rev.\ D {\bf 52}, 985 (1995)
  [hep-th/9502146].

%\cite{Candlish:2007fh}
\bibitem{Candlish:2007fh} 
  G.~N.~Candlish and H.~S.~Reall,
  %``On the smoothness of static multi-black hole solutions of higher-dimensional Einstein-Maxwell theory,''
  Class.\ Quant.\ Grav.\  {\bf 24}, 6025 (2007)
  [arXiv:0707.4420 [gr-qc]].






%\cite{Kubiznak:2006kt}
\bibitem{Kubiznak:2006kt} 
  D.~Kubiznak and V.~P.~Frolov,
  %``Hidden Symmetry of Higher Dimensional Kerr-NUT-AdS Spacetimes,''
  Class.\ Quant.\ Grav.\  {\bf 24}, F1 (2007)
  [gr-qc/0610144].
  %%CITATION = GR-QC/0610144;%%

%\cite{Page:2006ka}
\bibitem{Page:2006ka} 
  D.~N.~Page, D.~Kubiznak, M.~Vasudevan and P.~Krtous,
  %``Complete integrability of geodesic motion in general Kerr-NUT-AdS spacetimes,''
  Phys.\ Rev.\ Lett.\  {\bf 98}, 061102 (2007)
  [hep-th/0611083].
  %%CITATION = HEP-TH/0611083;%%

%\cite{Frolov:2006pe}
\bibitem{Frolov:2006pe} 
  V.~P.~Frolov, P.~Krtous and D.~Kubiznak,
  %``Separability of Hamilton-Jacobi and Klein-Gordon Equations in General Kerr-NUT-AdS Spacetimes,''
  JHEP {\bf 0702}, 005 (2007)
  [hep-th/0611245].
  %%CITATION = HEP-TH/0611245;%%


%\cite{Frolov:2003en}
\bibitem{Frolov:2003en} 
  V.~P.~Frolov and D.~Stojkovic,
  %``Particle and light motion in a space-time of a five-dimensional rotating black hole,''
  Phys.\ Rev.\ D {\bf 68}, 064011 (2003)
  [gr-qc/0301016].

%\cite{Gooding:2008tf}
\bibitem{Gooding:2008tf} 
  C.~Gooding and A.~V.~Frolov,
  %``Five-Dimensional Black Hole Capture Cross-Sections,''
  Phys.\ Rev.\ D {\bf 77}, 104026 (2008)
  [arXiv:0803.1031 [gr-qc]].

%\cite{Hackmann:2008tu}
\bibitem{Hackmann:2008tu} 
  E.~Hackmann, V.~Kagramanova, J.~Kunz and C.~L\"ammerzahl,
  %``Analytic solutions of the geodesic equation in higher dimensional static spherically symmetric space-times,''
  Phys.\ Rev.\ D {\bf 78}, 124018 (2008)
  [Erratum-ibid.\  {\bf 79}, 029901 (2009)]
  [arXiv:0812.2428 [gr-qc]].

%\cite{Kagramanova:2012hw}
\bibitem{Kagramanova:2012hw} 
  V.~Kagramanova and S.~Reimers,
  %``Analytic treatment of geodesics in five-dimensional Myers-Perry space--times,''
  Phys.\ Rev.\ D {\bf 86}, 084029 (2012)
  [arXiv:1208.3686 [gr-qc]].

%\cite{Hoskisson:2007zk}
\bibitem{Hoskisson:2007zk} 
  J.~Hoskisson,
  %``Particle Motion in the Rotating Black Ring Metric,''
  Phys.\ Rev.\ D {\bf 78}, 064039 (2008)
  [arXiv:0705.0117 [hep-th]].
  %%CITATION = ARXIV:0705.0117;%%

%\cite{Durkee:2008an}
\bibitem{Durkee:2008an} 
  M.~Durkee,
  %``Geodesics and Symmetries of Doubly-Spinning Black Rings,''
  Class.\ Quant.\ Grav.\  {\bf 26}, 085016 (2009)
  [arXiv:0812.0235 [gr-qc]].
  %%CITATION = ARXIV:0812.0235;%%

%\cite{Elvang:2006dd}
\bibitem{Elvang:2006dd} 
  H.~Elvang, R.~Emparan and A.~Virmani,
  %``Dynamics and stability of black rings,''
  JHEP {\bf 0612}, 074 (2006)
  [hep-th/0608076].
  %%CITATION = HEP-TH/0608076;%%

%\cite{Igata:2010ye}
\bibitem{Igata:2010ye}
  T.~Igata, H.~Ishihara and Y.~Takamori,
  %``Stable Bound Orbits around Black Rings,''
  Phys.\ Rev.\ D {\bf 82}, 101501 (2010)
  [arXiv:1006.3129 [hep-th]].
  %%CITATION = ARXIV:1006.3129;%%

%\cite{Armas:2010pw}
\bibitem{Armas:2010pw} 
  J.~Armas,
  %``Maximal Analytic Extension and Hidden Symmetries of the Dipole Black Ring,''
  Class.\ Quant.\ Grav.\  {\bf 28}, 235014 (2011)
  [arXiv:1011.5618 [hep-th]].
  %%CITATION = ARXIV:1011.5618;%%

%\cite{Grunau:2012ai}
\bibitem{Grunau:2012ai} 
  S.~Grunau, V.~Kagramanova, J.~Kunz and C.~L\"ammerzahl,
  %``Geodesic Motion in the Singly Spinning Black Ring Spacetime,''
  Phys.\ Rev.\ D {\bf 86}, 104002 (2012)
  [arXiv:1208.2548 [gr-qc]].
  %%CITATION = ARXIV:1208.2548;%%

%\cite{Grunau:2012ri}
\bibitem{Grunau:2012ri} 
  S.~Grunau, V.~Kagramanova and J.~Kunz,
  %``Geodesic Motion in the (Charged) Doubly Spinning Black Ring Spacetime,''
  Phys.\ Rev.\ D {\bf 87}, 044054 (2013)
  [arXiv:1212.0416 [gr-qc]].

%\cite{Igata:2013be}
\bibitem{Igata:2013be} 
  T.~Igata, H.~Ishihara and Y.~Takamori,
  %``Stable Bound Orbits of Massless Particles around a Black Ring,''
  Phys.\ Rev.\ D {\bf 87}, 104005 (2013)
  [arXiv:1302.0291 [hep-th]].

%\cite{Igata:2010cd}
\bibitem{Igata:2010cd}
  T.~Igata, H.~Ishihara and Y.~Takamori,
  %``Chaos in Geodesic Motion around a Black Ring,''
  Phys.\ Rev.\ D {\bf 83}, 047501 (2011)
  [arXiv:1012.5725 [hep-th]].



\end{thebibliography}
\end{document}